\documentclass[a4paper,11pt]{article}

\usepackage{cite}
\usepackage{graphicx}
\usepackage{epstopdf}
\usepackage{color}

\begin{document}

\title{Controllable emission of a dipolar source coupled with a magneto-dielectric resonant subwavelength scatterer}

\author{Brice Rolly$^1$, Jean-Michel Geffrin$^1$, Redha Abdeddaim$^1$, Brian Stout$^1$, Nicolas Bonod$^{1*}$}

\date{}

\maketitle

$^1$ CNRS, Aix-Marseille Universit\'e, Centrale Marseille, Institut Fresnel, UMR 7249, Campus de St J\'er\^ome, 13397 Marseille, France

$^{*}$ Corresponding author: nicolas.bonod@fresnel.fr
\vspace{1.5cm}

\textbf{
We demonstrate experimentally and theoretically that a local excitation of a single scatterer of relative dielectric permittivity $\varepsilon=6$ permits to excite broad dipolar and quadrupolar electric and magnetic resonances that shape the emission pattern in an unprecedented way. By suitably positioning the feed with respect to the sphere at a $\lambda/3$ distance, this compact antenna is able to spectrally sort the electromagnetic emission either in the forward or in the backward direction, together with a high gain in directivity. Materials with $\varepsilon=6$ can be found in the whole spectrum of frequencies promising Mie antennas to become an enabling technology in numbers of applications, ranging from quantum single photon sources to telecommunications.}\\

\section*{Introduction}
Mie resonances in spherical particles of moderate refractive index were recently shown to exhibit well pronounced electric and magnetic modes of both dipole and quadrupole orders~\cite{Evlyukhin10,Evlyukhin11,GarciaEtxarri11,Kuznetsov12,Evlyukhin12,Rolly12b,Schmidt12,Shi13}.
As first predicted by Kerker in the case of particles with magnetic permeability~\cite{Kerker1983}, interference effects between electric and magnetic modes can strongly favour either forward or backward scattering, respectively depending on the frequency of the incident plane wave~\cite{Gomezmedina11,Nieto-Vesperinas11,Liu12,Geffrin12,Person12,Fu13} with applications in anti-reflecting structured surfaces for photovoltaic cells~\cite{Spinelli12}. Dielectric particles reveal to be highly interesting scatterers to design highly directive, compact and lossless antennas~\cite{Devilez10,Krasnok11,Filonov12,Krasnok12}. Recently, it was proposed to extend the so-called Kerker's conditions to the case of near field excitations, where the interplay between electric and magnetic modes can boost the directivity of antennas~\cite{Rolly12c}. It was also proposed to realize a notch inside a dielectric scatterer of refractive index $n$ around 4 in order to increase the number of multipoles excited by a near field source~\cite{krasnok12b}. Such magneto-dielectric antennas offer more parameters for optimizing the emission directivity than classical metallic elements which behave principally as electric multipolar scatterers~\cite{Li07,Curto10,Esteban10,Bonod10,Rolly11b,Massa13} and they could be very interesting to direct the fluorescence emission of quantum emitters in the optical range of frequencies. 

Here we propose to design a versatile antenna with the use of a single spherical scatterer exhibiting a refractive index $n=2.45$. Mie resonators with refractive index typically around 3.5 have been mainly considered because they exhibit narrow and well pronounced electric and magnetic modes. A decrease of the refractive index leads to a broadening of the Mie resonances and a refractive index $n=2.45$ allows for a remarkable mixing between electric and magnetic dipolar and quadrupolar modes over a finite range of frequencies. We show that the interplay between the first two electric and magnetic modes permits to control the emission directivity by tuning the emission frequency. Also, lessening the refractive index opens the way towards the use of oxides such as rutile TiO$_{2}$ to design highly directive optical antennas. 

\section*{Results}
We measure the emission pattern in the E-plane (\textit{i.e.} $xOz$ plane, see Fig.1d) of a subwavelength scale `Mie' antenna consisting of a single dielectric particle of permittivity $\varepsilon=6$ ($n$=2.45) fed in the GHz regime by a two-arm electric dipole emitter. By controlling the emitter-to-particle distance at a deep subwavelength scale, we report on the possibility to choose the emission direction by tuning the frequency from 8.7 GHz to 9.74 GHz. Experiments are carried out in an anechoic chamber (CCRM-Marseille, see Fig.1a) dedicated to amplitude and phase measurements of the electric field with a receiver antenna rotating along a circular arm 4 m in diameter, centered on the feed~\cite{Geffrin12}. Experimental measurements are accompanied by a theoretical derivation of the emission pattern based on the coupling between the feed and the magnetic and electric modes of the sphere. 

\begin{figure}[!htb]
\begin{center}
\includegraphics[width=0.5\linewidth]{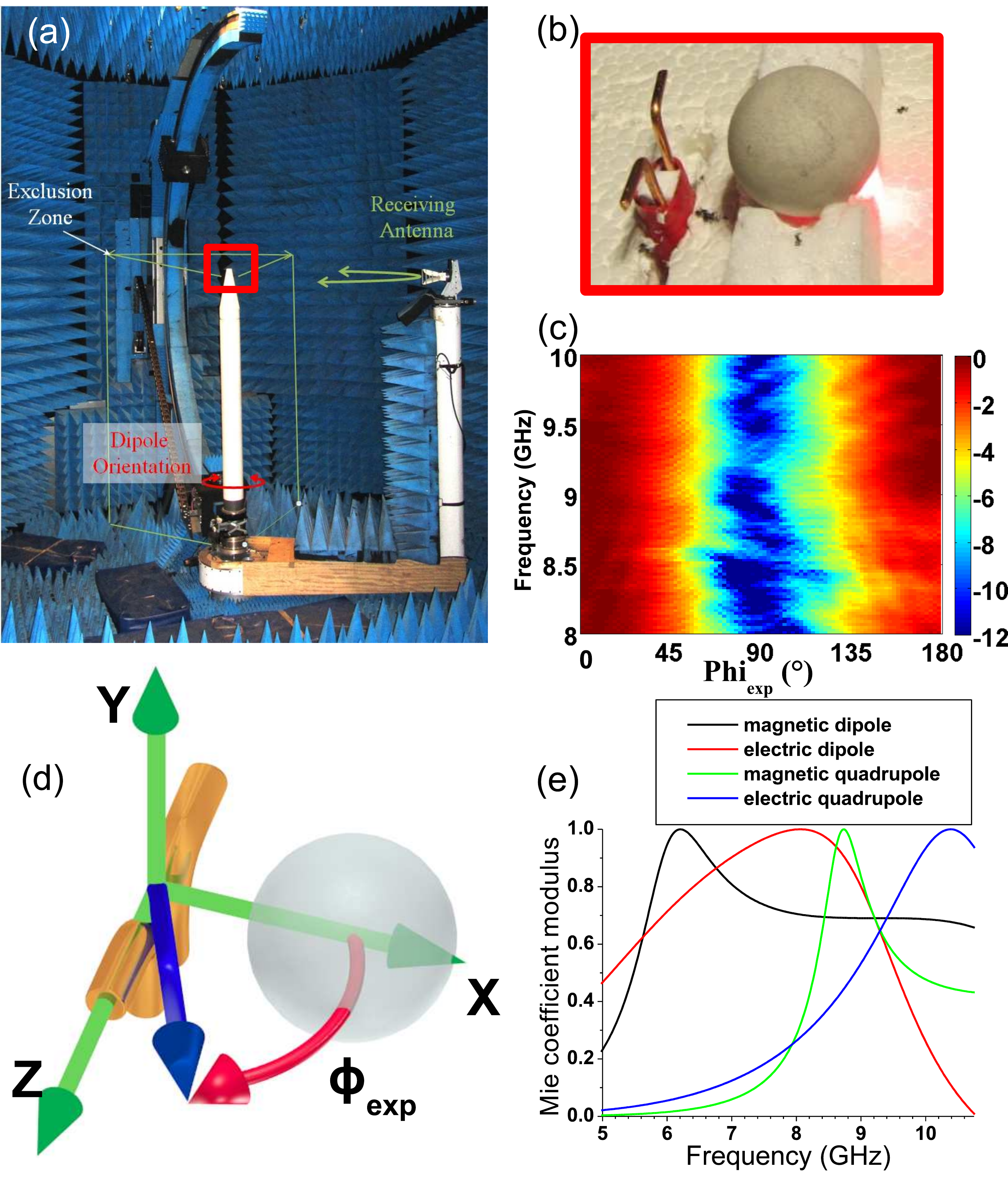} 
\end{center}
\caption {(a) Photograph of the experiment in the anechoic chamber: the dipole-sphere antenna is placed on a polystyrene mast at the center of an anechoic chamber. The receiver antenna (ARA DRG 118) can rotate around the $Oy$ axis (see d) except in an exclusion zone in the angle range $[-130^\circ, 130^\circ]$ due to the presence of the vertical arch. (b) Close-up of the dipole-sphere antenna and of the polystyrene holders. (c) Norm of the electric field emitted by the isolated dipole antenna as a function of the frequency and angle $\phi_{\rm exp}$ in the E-plane ($xOz$ plane) in dB. At each frequency, the field is normalized by its maximum in the E-plane. (d) Sketch of the experiment with the axes and $\phi_{\rm exp}$ angle in the $(xOz)$ plane, used in Fig.1c, Fig.2 and Fig.3; the blue arrow indicates the direction of the receiving antenna. (e) Norm of the Mie coefficients $|c_{j}^\mathrm{e(m)}|$ with respect to frequency in GHz. Black line: magnetic dipole $|c_{1}^\mathrm{m}|$; red line: electric dipole $|c_{1}^\mathrm{e}|$; green line: magnetic quadrupole $|c_{2}^\mathrm{m}|$; blue line: electric quadrupole $|c_{2}^\mathrm{e}|$.}
\end{figure}

The two-wire dipole is directly attached to a connector (SubMiniature version A). Each arm is 9 mm long which results in a total length smaller than the wavelength even at the highest operating frequency.
The characterization of the dipole in its E-plane (Fig. 1c) shows a classical dipole-like radiation pattern, with comparable front and backward  radiated amplitude (respectively $0^\circ$ and $180^\circ$), and negligible emission in the $90^\circ$ direction.
The emitter is coupled with a single dielectric sphere, 19 mm in diameter, composed of an Eccostock HIK (Emmerson $\&$ Cumming) material of permittivity 6, presenting low losses ($\tan\delta\le 0.02$). The distance between the dipole and the sphere can be controlled at a submillimeter scale via an expanded polystyrene holder equivalent to air at the operating frequencies (see Fig.1a-b). Measurements have been performed with emitter-to-sphere distances of 5, 10 and 20 mm and frequencies ranging from 8 to 10 GHz with a step of 20 MHz, which corresponds to a total of 300 emission pattern measurements (the emitter-to-sphere distance is defined between the emitter and the surface of the sphere, i.e. it is equal to emitter-to-sphere center distance minus the sphere radius). 
The amplitude of the electric field is displayed in Fig. 2 with respect to the frequency and angle $\phi_{\rm exp}$ ranging from 0 to 180$^\circ$, i.e. in a half-plane containing one of the emitter arms. At every frequency, the amplitude is normalized by the maximum measured in the [0;180$^\circ$] range, and is displayed in dB. One observes that the radiation pattern is highly sensitive to modifications of the emitter-to-particle distance on a scale much smaller than the emission wavelength. Importantly, we observe that this Mie antenna can emit either in the forward (around 8.75 GHz) or backward (around 9.5 GHz) direction efficiently at a 10 mm distance. \\
In order to demonstrate the contribution of each mode in tuning the emission directivity, we derived a formula of the irradiance in the E-plane (derivation is detailed in the Supplementary information). 
Such a formula invokes the far field expression of the interference between the fields produced by (i) the dipolar source and (ii) the induced electric-magnetic dipoles and quadrupoles in the dielectric particle.
The fields are derived in the spherical vector basis,  $[\hat{\mathbf{e}}_\mathrm{r},
\hat{\mathbf{e}}_\mathrm{\theta},\hat{\mathbf{e}}_\mathrm{\phi}]$ with the dipole emitter defined to lie along the $+z$ axis ($\mathbf{p}_0\equiv p_0 \hat{\mathbf{z}}$) and placed at the origin of the coordinate system. The center of the spherical scatterer is placed at a distance $d$ from the emitter on the $x$-axis, $\mathbf{r}_1=+d\hat{\mathbf{x}}$.
The total, normalized (respectively to the electric dipole emitter maximum far-field irradiance) far-field irradiance can be cast in the standard spherical coordinate system:
\begin{eqnarray}
I(\theta,\phi) &=&|\tilde{\mathbf{E}}(\mathbf{r})|^2  \nonumber \\
&=& \bigg| \sin\theta \hat{\mathbf{e}_\theta}  \nonumber \\ 
&& + e^{i\varphi} \gamma^e_1 \tilde{\alpha}^e_1 \sin\theta\hat{\mathbf{e}_\theta} \nonumber  \\
&& + e^{i\varphi}\gamma^m_1 \tilde{\alpha}^m_1 (\cos\phi\hat{\mathbf{e}_\theta} - \sin\phi\cos\theta \hat{\mathbf{e}_\phi} )  \nonumber \\
&& + e^{i\varphi}\gamma^e_2 \tilde{\alpha}^e_2 ( \cos\phi\cos2\theta\hat{\mathbf{e}_\theta} -\sin\phi\cos\theta\hat{\mathbf{e}_\phi}) \nonumber \\
&& + e^{i\varphi}\gamma^m_2 \tilde{\alpha}^m_2 ( \cos2\phi\sin\theta\hat{\mathbf{e}_\theta} -
\frac{\sin2\phi\sin2\theta}{2}\hat{\mathbf{e}_\phi})\bigg|^2 
\end{eqnarray} 
In the latter expression, each subsequent line stands for the field produced by the source and the induced electric dipole, magnetic dipole, electric quadrupole, and magnetic quadrupole respectively; $e^{i\varphi}$ is a far-field phase shift contribution; $\tilde{\alpha}_n^{e(m)}$ is a \emph{dimensionless} electric (magnetic) polarizability of order $n$, defined in terms of Mie theory and a T-matrix formalism (see Supporting Information). The first four Mie coefficients are plotted in Fig.1e with respect to the frequency and one remarks that the relative low refractive index of the resonant scatterer leads to broad resonances and that the norms of the four Mie coefficients are nearly equal near 9.75 GHz. The coupling coefficients between the emitter and the first 2 electric and magnetic modes of the sphere can be cast \cite{Rolly11}:
\begin{eqnarray*}
\gamma^e_1 &\equiv &  e^{ikd} (k^2d^2+ikd-1)(a/d)^3    \\
\gamma^m_1 &\equiv & e^{ikd} (k^2d^2+ikd)(a/d)^3    \\
\gamma^e_2 &\equiv & - \frac{5}{3} e^{ikd} (k^3d^3 +3 i k^2d^2 - 6 kd - 6 i) (a/d)^4    \\
\gamma^m_2 &\equiv &   \frac{5}{3} e^{ikd} (k^3d^3 + 3ik^2d^2-3kd) (a/d)^4   \\
\end{eqnarray*}

In the $(xOz)$ ($\phi=0$ or $\pi$) plane, the irradiance can be cast:
\begin{eqnarray}
I^{\pm xz}(\theta) &=& \bigg| \sin\theta (1 + e^{-ikd\pm\sin\theta} \gamma^e_1 \tilde{\alpha}^e_1) \nonumber \\
&\ & - e^{-ikd\pm\sin\theta} \times \big( \pm\gamma^m_1 \tilde{\alpha}^m_1  \pm  \gamma^e_2 \tilde{\alpha}^e_2 \cos2\theta \nonumber \\
 &\ & + \gamma^m_2 \tilde{\alpha}^m_2 \sin\theta \bigg|^2 
\end{eqnarray} 

\begin{figure}[!htb]
\begin{center}
\includegraphics[width=0.9\linewidth]{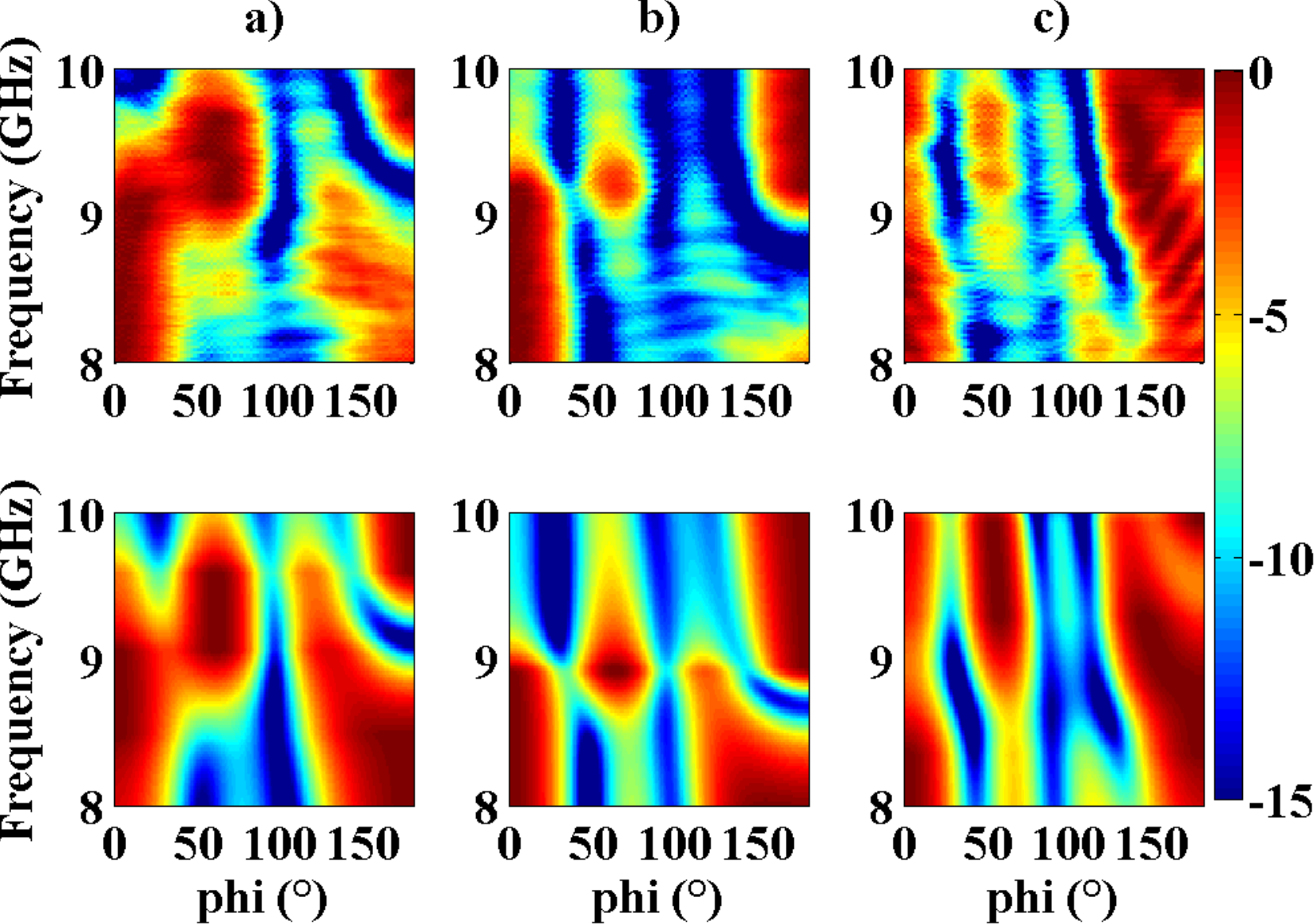} 
\end{center}
\caption{Norm of the electric field in the E-plane, in decibels, measured (top row) and simulated (bottom row) as functions of the antenna receiver angle $\phi_{\rm exp}$ in degrees (in abscissa) and emitting frequency in GHz (in ordinate). At each frequency, the field is normalized by its maximum in the E-plane. The emitter-to-sphere gaps are (a) 5 mm, (b) 10 mm and (c) 20 mm.} \end{figure}

The normalized amplitude of the electric field in the E-plane is displayed in Fig.2 as a function of the angle in
the $(xOz)$ plane, and a good agreement between theory and experiments is observed for the three 
emitter-to-particle distances despite a small frequency shift that is likely due to an imperfect knowledge of the
permittivity of the sphere (the real part is given at $\pm{5}\%$ and the imaginary part is set to 0 in the model) and to the
physical length of the source, the theoretical expression assuming a point dipole. This model is general, and could 
also be applied to predict the scattering pattern of metallic magneto-electric antennas provided that the diagonal
polarizability tensor elements are accurately determined and predominate over the extra-diagonal elements. 

\section*{Discussion}
Both theoretical model and experiments predict that the privileged direction of emission of the antenna can be controlled for a 10 mm separation distance by tuning the emission frequency. This feature is highlighted in Figs.3a and b where the intensities of the scattered electric field measured in the E-plane at frequencies 8.7 GHz and 9.74 GHz are plotted in polar coordinates. Measurements and theory are quantified by defining directivity gain, in isotropic decibels, $D_\mathrm{dBi}=10\log\big( 4\pi I / P_\mathrm{rad} \big)$ where $I$ and $P_\mathrm{rad}$ are respectively the radiant intensity in the direction of interest and the total radiative power of the antenna. Theoretical values of 7.01 dBi at 8.66 GHz and 5.17 dBi at 9.58 GHz in the forward and backward directions respectively can be achieved. The sphere 	also effectively increases antenna gain: the presence of the sphere increases the measured electric field intensity: by a factor of 3.4 and 2.4 along the corresponding privileged directions at 8.7 GHz and 9.74 GHz respectively.

\begin{figure}[!htb]
\begin{center}
\includegraphics[width=0.8\linewidth]{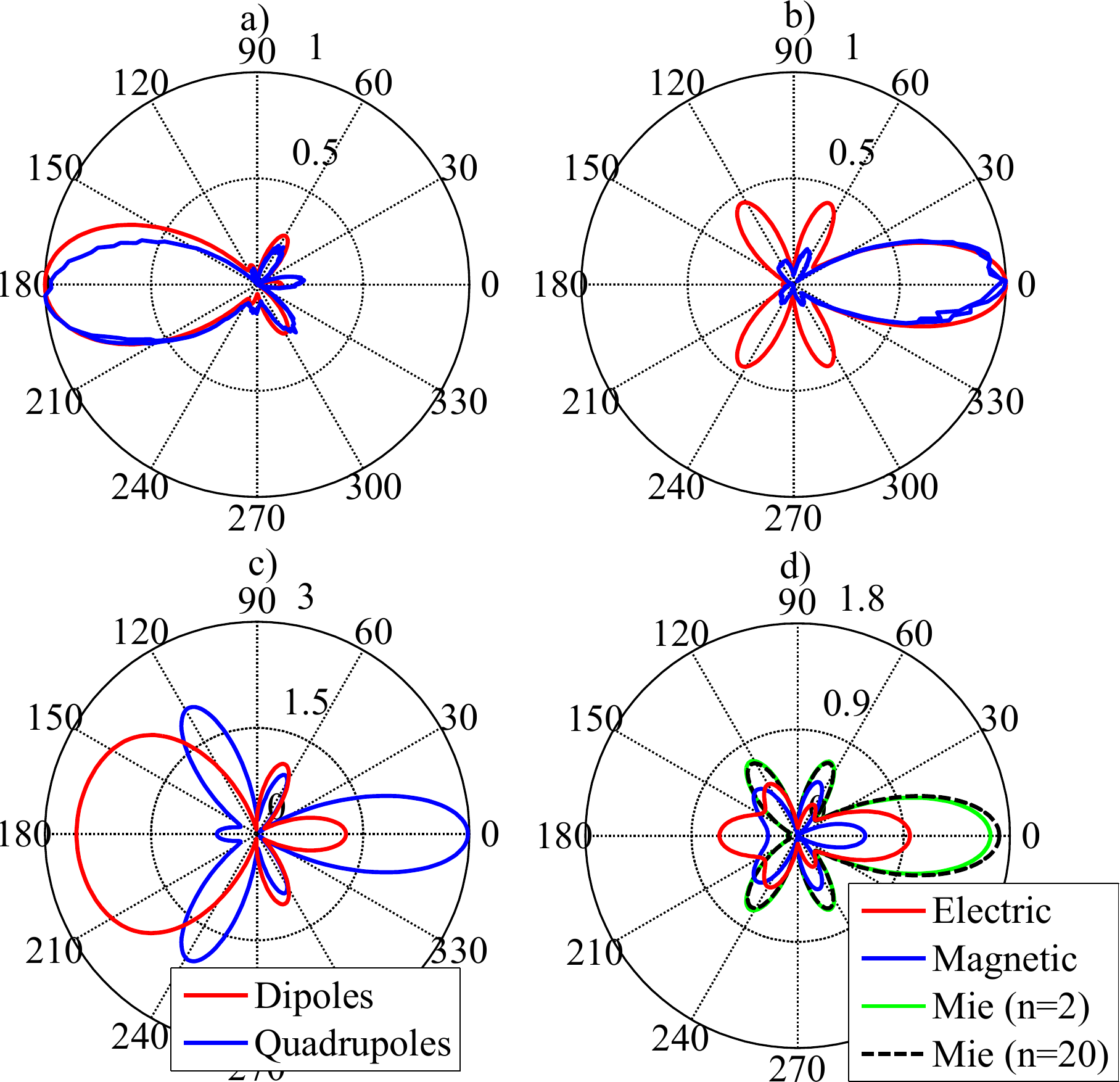} 
\end{center}
\caption{(a,b) Norm of the electric field scattered for an emitter-to-particle gap of 10 mm in the E-plane, (red line) simulations with 20 multipoles and (blue) measurements: (a) Back-scattering observed at 9.74 GHz, (b) Forward scattering observed at 8.7 GHz. (c) Emission pattern derived with Eq. 2 when considering the induced dipoles (red line) or quadrupoles (blue line) only at 8.7 GHz. (d) Emission pattern obtained with Eq. 2 when considering electric (red line) or magnetic modes (blue line) of dipolar and quadrupolar orders, sum of electric and magnetic response of both dipole and quadrupole order (green line), electric and magnetic modes with 20 multipoles (dashed black line) at 8.7 GHz.} 
\end{figure}


The emission patterns plotted in Fig.3c at $f$=8.7 GHz when considering dipoles or quadrupoles only (full red and blue lines) show that neither dipole nor quadrupole excitations taken alone suffice to explain the directivity: a purely response of the sphere would predict light preferentially emitted in the backward direction, whereas a solely quadrupole response would lead to forward emission together with a low gain in directivity. A similar filtering with respect to either purely  electric or magnetic modes, illustrated in Fig.3d, shows that neither purely electric nor purely magnetic modes can fit the emission pattern measured in the E-plane, while the emission pattern is accurately reproduced when both dipoles and quadrupoles are considered. Comparison with a multipolar calculation performed with a Generalized Mie theory calculation taking into account 20 multipole orders reveals the high accuracy of the quadrupolar model. These last two plots demonstrate that the emission pattern results from an efficient coupling between the electric and magnetic modes, of both dipolar and quadrupolar orders which is made possible by the broadness of the Mie resonances displayed in Fig.1e. The high sensitivity of the emission pattern with respect to the emission frequency is explained by a strong modulation of three of the relevant modes between 8.5 GHz and 10 GHz (see Fig.1e).

In conclusion, a single dielectric particle of refractive index $n=2.45$ is used to efficiently tune the scattering directivity of an electric dipole emitter. The tuning of the emission frequency modulates the amplitude and phase of the first four modes of the sphere which drastically modifies the emission pattern. Dielectric particles of refractive index around 2.5 can be found over a broad spectrum making dielectric Mie antennas interesting from microwaves to nano-optics where they could acts as lossless subwavelength spectral sorters~\cite{Aouani11,Shegai11}. 

\section*{Methods}

\subsection*{Measurements} 
The measurements are carried out in an anechoic chamber (Centre Commun de Ressources en Microondes in
 Marseille, France) that allows electric field amplitude and phase measurements thanks to an antenna receiver
placed on a circular arm of radius 2 m. For this study, the electric field is quantified by measuring the electric field component lying inside the E-plane (the electric field component normal to the E-plane being negligible). 
The $S_{21}$ parameter is measured to provide the norm of the electric field $20\log|E|$. 
The field is normalized at each frequency by the maximum value $E_{max}$ measured when varying the angle $\phi_{\rm exp}$ and the value of $20\log|E/E_{max}|$ is reported in Fig. 2. In Fig. 3, the field is also normalized by its maximum when varying the angle $\phi_{\rm exp}$ at frequencies 8.7 GHz and 9.74 GHz. 

The antenna receiver cannot make a full rotation around the axis due to the presence of the vertical arch. The exclusion zone is almost $100^{\circ}$. The complete $360^{\circ}$ emission pattern requires a second measurement for which the mast holder is rotated by $120^{\circ}$. In Figs.3a and b, it can be observed that the two measurements match and permit the reconstruction of the emission patterns with a high accuracy.
 
\subsection*{Theoretical calculations}
The far field expressions of the electric fields radiated by the feeding electric dipole, and the electric and magnetic dipoles and quadrupoles excited in the particle are derived and summed up to obtain the final expression of the total electric field. The sources are strongly coupled in near field and the calculation of the phase differences between the coherent fields must take into account the entire field, \textit{i.e.} near, intermediate and far field expressions. The expression of the emission pattern requires the calculation of the dipolar and quadrupolar polarizabilities of the scatterer, that can be easily obtained within the context of generalized Mie theory \cite{Stout11}. The analytical derivation is detailed in the Supporting Information. Numerical simulations are performed using an in-house Generalized Mie Theory code. The embedding medium is air and its refractive index is taken equal to 1, the dielectric permittivity of the spherical scatterer is estimated by Emmerson Cumming to be equal to $6\pm{5\%}$.

\section*{Acknowledgements}
The authors thank S\'ebastien Bidault for discussions and Jean-Pierre Spinelli for his involvement in the anechoic chambers. This work is supported by Agence Nationale de la Recherche via project ANR 11 BS10 002 02 TWINS.   

\section*{Author contributions} 
N.B, R.A. and J.M.G. conceived the experiment; R. A. and J.M.G. carried out the experiment; B. R. derived the analytical expression, performed the numerical calculations with B. S. and N. B.; All the authors contributed to the discussion and analysis of the results; N. B. wrote the manuscript with contribution from all the authors.

\section*{Additional information}
The authors declare no competing financial interests. Supplementary information accompanies this paper and include the detailed derivation of expression 2. Reprints and permissions information is available at www.nature.com/reprints. Correspondence and requests for materials should be addressed to N. B.

\end{document}